# A Novel DDoS Attack Detection Method Using Optimized Generalized Multiple Kernel Learning

Jieren Cheng[1,2], Junqi Li[2,*], Xiangyan Tang[2], Victor S. Sheng[3], Chen Zhang[2] and Mengyang Li[2]

**Abstract:** Distributed Denial of Service (DDoS) attack has become one of the most destructive network attacks which can pose a mortal threat to Internet security. Existing detection methods can not effectively detect early attacks. In this paper, we propose a detection method of DDoS attacks based on generalized multiple kernel learning (GMKL) combining with the constructed parameter R. The super-fusion feature value (SFV) and comprehensive degree of feature (CDF) are defined to describe the characteristic of attack flow and normal flow. A method for calculating R based on SFV and CDF is proposed to select the combination of kernel function and regularization paradigm. A DDoS attack detection classifier is generated by using the trained GMKL model with R parameter. The experimental results show that kernel function and regularization parameter selection method based on R parameter reduce the randomness of parameter selection and the error of model detection, and the proposed method can effectively detect DDoS attacks in complex environments with higher detection rate and lower error rate.

## 1 Introduction

A distributed denial of service (DDoS) attack is a network attack that causes bandwidth overload due to the use of traffic on the network, resulting in normal services not working properly. In recent years, attacks on broadcast systems, financial systems and Internet-based services have grown geometrically [Lee, Baik, Kim et al. (2018)]. Furthermore, such attacks are devastating, wide-ranging, easy to implement, and difficult to track and prevent, posing a major threat to the Internet security and may even threaten the natural ecosystems [Ye, Cheng, Zhu et al. (2018)]. The Internet has long been plagued by DDoS attacks. Especially in recent years, DDoS attacks have escalated dramatically. For example, attacks on DynDNS services cut off common Internet services such as Amazon and GitHub which occurred on October 21, 2016. According to a Radware survey, DDoS was currently the biggest threat for organizations (50% of respondents in the survey) [Li, Wu, Yuan et al. (2018)]. According to the World Infrastructure Security Report, the largest attack reported by a respondent in 2016 was 500 Gbps, with attacks reported by others were 450 Gbps, 425 Gbps, and 337 Gbps. According to the Abor network, which provides DDoS defense services, they have observed more than 124,000 DDoS attacks per week since 2016, and they believe that this number will have been growing rapidly [Bisson (2016); Yadav, Trivedi and Mehtre (2016)]. Different from other network attacks, DDoS attacks only need a large number of zombies and a small amount of cyber security knowledge then an effective attack can be lauched. [Wang, Zheng and Li (2017)]. This easy-to-follow network attack approach further enhances the power of DDoS attacks. Therefore, a timely and correct identification of DDoS attacks is imminent.

[1] Key Laboratory of Internet Information Retrieval of Hainan Province, Hainan University, Haikou 570228, China.

[2] College of Information Science & Technology, Hainan University, Haikou 570228, China.

[3] Department of Computer Science, University of Central Arkansas, Conway, AR 72035, USA

* Corresponding Author: Junqi Li. Email: junqi@hainanu.edu.cn.

Attack detection is one of the main DDoS defense mechanisms. However, in most cases, attack traffic is very similar to legitimate traffic. Attackers often use this feature to launch DDoS attacks. In most cases, low-rate and low-traffic attacks are often mistaken for legitimate activities at very early stages. Many researchers try to use different methods to detect DDoS attacks, but these methods still have some shortcomings. The collected data is characterized by high latitude and variable samples; the correlation between features is ignored, and some information is lost; the selection of kernel function is not the perfect theoretical basis, and the kernel function and regularization are not perfect. The selection of parameters is random, and manual optimization may be difficult, and the wrong selection of parameters will affect the accuracy of DDoS attack detection.

In this paper, a method of DDoS detection based on generalized multiple kernel learning (GMKL) is proposed, and the above problems and shortcomings are analyzed and improved. This method consists of two stages: (1) extracting the traffic characteristics of normal networks, defining the superfusion feature value (SFV) and the comprehensive degree of feature (CDF) based on in-depth analysis of the traffic characteristics of normal networks; (2) to solve the problem of parameters selection, we propose the R parameter for kernel functions and regularization parameters. Based on the proposed features, an improved GMKL method is used to detect DDoS attack. The experimental results show that compared with support vector machine (SVM) and simple multiple kernel learning (SMKL) methods, our method can more accurately distinguish between normal flow and attack flow.

The remainder of this paper is structured as follows. We describe related work in Sec.2. In Sec.3, we present a detailed description of our proposed method, including the feature selection, the architecture of the algorithm and the method to select the combination of kernel function and regularization paradigm. Experimental setup and results are presented in Sec.4. This section also gives the results of comparison with other different methods. Eventually, we draw a conclusion on our method in Sec.5.

**2 Related work**

With the development of DDoS attack intrusion detection technology, current research of DDoS attack detecting is still grim. There are still high misdiagnosis rate and omissive judgment rate in network traffic processing and new DDoS attack detection. Nowadays, summing up the researches on detecting technology for DDoS attack, they are usually implemented based on different network environment. According to the network environment, these methods can be divided into three categories: detection methods in the conventional network environment, the cloud environment, and the software-defined network (SDN) environment.

(1) A conventional network environment refers to an open system interconnection (OSI) based Internet environment commonly used on the Internet today. A novel method based on k-nearest neighbor traffic classification and correlation analysis which was used as a way to detect DDoS attack was proposed by Xiao et al. [Xiao, Qu, Qi et al. (2015)]. By utilizing the entropy difference between business streams, a universal detection algorithm is proposed by Behal et al. [Behal and Kumar (2017)] which could detect different types of DDoS attacks. A framework was proposed by Han et al. [Han, Bi, Liu et al. (2017)] which was based on Spark's new DDoS attack detection system including information entropy-based algorithm and dynamic sampling k-means parallel algorithm. The method proposed by Hoque et al. [Hoque, Kashyap and Bhattacharyya (2017)] is a real-time detection method for DDoS which can identify DDoS attack and generate high detection accuracy. Chen et al. [Chen, Wu, Ye et al. (2013)] proposed a new detection model based on conditional random field (CRF) which combined signature based and anomaly based detection methods into a hybrid system.

(2) The development of cloud computing and the maturity of cloud technology have gradually increased the importance of cloud security issues and become an important factor restricting the development of cloud computing. DDOS detection is required in fields such as cloud robots [Liu, Wang, Liu et al. (2019)]. A dynamic resource allocation mechanism is proposed by Yu et al. [Yu, Tian, Guo et al. (2014)] which is based on the dynamic allocation of idle resources from the cloud to the victim machine, and that will acquire the quality of service. A method of filtering out high-data packets proposed by Preeti et al. [Daffu and Kaur (2016)] can avoid DDoS attack in the cloud environment. Another method to detect DDoS attack is proposed by Himadri et al. [Mondal, Hasan, Hossain et al. (2017)] using fuzzy logic in the cloud environment. Karnwal et al. [Karnwal, Sivakumar and Aghila (2012)] proposes the DDoS attack defense methods under the cloud computing platform XML and HTTP. A novel collaborative solution framework proposed by Somani et al. [Somani, Gaur, Sanghi et al. (2017)] is based on multi-level alert flows for DDoS attack of cloud services.

(3) SDN separates the control plane of the network from the data forwarding plane, and it can allocate and schedule network resources flexibly and efficiently. It has been widely used in the network field, but it also brings new security problems. Researchers have also done a lot of research on its network security. An architecture proposed by Gharakheili et al. [Gharakheili, Bass, Exton et al. (2014)] includes a cloud-based front-end user interface and a back-end SDN-based API that allows ISPs to permit users to customize their home network experience. A method for analyzing the behavior of DDOS attacks under SDN is proposed by N. Dayal et al. [Dayal and Srivastava (2017)], and the detection characteristics of DDOS are defined. Seeber et al. [Seeber and Rodosek (2014)] believe that SDN provides a new opportunity for network security in cloud scenario, because SDN based cloud provides greater flexibility and faster response when conditions change. An intelligent elastic risk assessment method proposed by Mihai-gabriel et al. [Mihai-Gabriel and Victor-Valeriu (2014)] is based on neural network and risk theory in SDN environment.

Although the above methods can extract representative features and have high detection accuracy in their specific application environments, the correlation between features is neglected and part of the information is lost [Cheng, Yin, Liu et al. (2009a,b); Cheng, Tang and Yin (2017); Cheng, Liu and Tang (2018)]. In recent years, there are some problems of feature matching and feature selection in many fields [Li, Qin, Xiang et al. (2018); Cai, Wang, Zheng et al. (2013); Fang, Cai, Sun et al. (2018).]. For attack detection, the importance of different features is also different. When multiple features are used to describe the information of DDoS attack, if only multiple simple combinations are combined into one feature vector, there are obvious disadvantages: the coefficient of each feature cannot be adjusted, which to some extent affects the later detection efficiency. Moreover, when the sample features involve a large scale and the spatial distribution of multi-dimensional data is uneven in high-dimensional feature space, using the conventional single-kernel method to process the sample cannot achieve the ideal detection effect [Petkovic, Basicevic, Kukolj et al. (2018)]. Most importantly, there is no perfect theoretical basis for the selection of kernels. The selection of kernels and regularization parameters is random. Manual optimization may be difficult, because it is also difficult to select and combine proper kernels and regularization paradigms [Cheng, Xu, Tang et al. (2018)].

In order to solve the above problems, we propose a new DDoS attack detection method. By defining two complementary eigenvectors, i.e. SFV and CDF, and then propose R parameter for the selection of kernel function and regularization parameter. Finally, based on the proposed feature, we adopt an improved GMKL algorithm. By comparing with other experimental methods, our method has greatly improved the recognition rate of normal flow and attack flow.

## 3 DDoS attack detection model based on R-GMKL

### *3.1 Feature extraction*

After analyzing the characteristics of the attack stream, based on our previous work [Cheng, Zhang, Tang et al. (2018)], we made the following definitions. Assume that network flow T in a certain unit of time F is $<(t_1,s_1,d_1,p_1),(t_2,s_2,d_2,p_2),......,(t_n,s_n,d_n,p_n)>$, where $t_i$、$s_i$、$d_i$、$p_i$ denotes the time, source IP address, destination IP address, and port number of the $i(i=1,2,.....,n)$-th data packet respectively. The class with all IP address having IP address as $A_i$ and destination IP address $A_j$ is $SD(A_i,A_j)$, the class with all data packets with source IP address as $A_i$ are $IPS(A_i)$, and all data packets with destination IP address as $A_j$ are $IPD(A_j)$. Remember that the source IP address $IPS(A_i)$ makes the class $IPS(A_i)$ and class $IPD(A_i)$ non-null packets to be $IF(A_i)$; note that the source IP address $IPS(A_i)$ that makes class $IPD(A_i)$ empty in class $A_i$ is $SH(A_i)$, and the number of different port numbers in $SH(A_i)$ is denoted as $Port(SH(A_i))$. In class $IPD(A_i)$, the destination IP address $A_i$ that is null in class $IPS(A_i)$ is $DH(A_i)$, and the number of different port numbers in $DH(A_i)$ is denoted as $Port(DH(A_i))$.

**Definition 1.** If there are different destination IP addresses $A_j$ and $A_k$, making classes $SD(A_i, A_j)$ and $SD(A_i, A_k)$ are not empty, then delete the class where all source IP address $A_i$ packets reside.

Assume that the last remaining class $ACS_1, ACS_2, ....., ACS_m$, is the address correlation degree that defines the network flow.

$$ACD_F = \sum_{i=1}^{m} W(ACS_i) \tag{1}$$

where $W(ACS_i) = \theta_1 Port(ACS_i) + (1-\theta_1) Packet(ACS_i), (0 < \theta_1 < 1)$, $Port(ACS_i)$ is the number of different port numbers in class $ACS_i$, $Packet(ACS_i)$ is the number of data packets in class $ACS_i$, and $\theta_1$ is the weighted value.

**Definition 2.** If all the packets whose destination IP address is $A_j$ form only a unique class $SD(A_i, A_j)$, delete the class where the packet with the destination IP address is $A_j$.

Assuming that the last remaining class is $SDS_1, SDS_2, ....., SDS_l$, classify the $l$ classes, that is, classify the data packets of the same class with the same destination IP address into the same class. The class of all data packets with the IP address of $A_j$ is $SDD(Aj)$. The class $SDD_1, SDD_2, ....., SDD_m$ is defined as the IP Flow Features Value (IP Flow Characteristics Value) of the network flow.

$$FFV_F = (\sum_{i=1}^{m} CIP(SDD_i) - m) \tag{2}$$

$CIP(SDD_i)$ in eq.2 are counted by:

$$CIP(SDD_i) = Num(SDD_i) + \theta_2 \sum_{j=1}^{Num(SDD_i)} OA(Pack(A_j)) + (1-\theta_2)OB(Port(SDD_i) - 1)$$

$$(0 \leq \theta_2 \leq 1) \tag{3}$$

where $Num(SDD_i)$ is the number of different source IP in $SDD_i$,

$$OA(Pack(A_j)) = \begin{cases} Pack(A_j) & Pack(A_j)/\Delta t > \theta_3 \\ 0 & Pack(A_j)/\Delta t \leq \theta_3 \end{cases}, Pack(A_j) \text{ is the number where IP}$$

equals $A_j$ in $SDD_i$, $\theta_3$ is threshold,

$$OB(Port(SDD_i)) = \begin{cases} Port(SDD_i) & Port(SDD_i)/\Delta t > \theta_4 \\ 0 & Port(SDD_i)/\Delta t \leq \theta_4 \end{cases}, Port(SDD_i) \text{ is the}$$

number of different target port in $SDD_i$, $\theta_4$ is threshold, $\Delta t$ is sampling interval.

**Definition 3.** Assume the IF flow as $IF_1, IF_2, ..., IF_M$, SH flow as $SH_1, SH_2, ..., SH_S$, DH flow as $DH_1, DH_2, ..., DH_D$, then define IP Flow Interaction Behavior Feature, $IBF$ as

$$IBF = \frac{1}{M+1}(|S-D| + \sum_{i=1}^{S} over(Port(SH_i)) + \sum_{i=1}^{D} over(Port(DH_i))) \quad (4)$$

$$over(x) = \begin{cases} x & x/\Delta t > \theta_5 \\ 0 & x/\Delta t \leq \theta_5 \end{cases}, \text{ where } \theta_5 \text{ are threshold. } M \text{ in eq.4 means the total IF flow}$$

in OP within $\Delta t$, |S-D| means the absolute value of the difference between the number of source IP addresses and the number of destination IP addresses for all HF flows in $\Delta t$.

**Definition 4.** Assume that the resulting SD classes are $SD_1, SD_2, ..., SD_L$, and IF classes are $IF_1, IF_2, ..., IF_L$. The number of packets of source IP address $A_i$ in class $IF_i$ is recorded as $SN_i$, where $i = 1, 2, ..., M$, the number of packets of all interworking flow classes is denoted as SN, and the source semi-interactive flow class is $SH_1, SH_2, ..., SH_S$, The number of different port numbers in class $SH_i$ is denoted as $Port(DH_i)$, where $i = 1, 2, ..., S$. The destination semi-interactive class is $DH_1, DH_2, ..., DH_D$, the number of different port numbers in class $DH_i$ is denoted as $Port(DH_i)$, where $i = 1, 2, ..., D$.

The weighted value of the abnormal number of all SH class packets is

$$Weight_{SH} = \sum_{i=1}^{S} oversh(Packet(SH_i)) \quad (5)$$

The anomalous weighted number of all SD class packets is

$$Weight_{SD} = \sum_{i=1}^{L} oversd(Packet(SD_i)) \quad (6)$$

The abnormal weighted value of the number of packets of network flow $F$ in unit time $T$ is

$$Weight_{packet} = flag(Weight_{SD})Weight_{SD} + Weight_{SD} \quad (7)$$

where $oversh(x) = \begin{cases} x, x/\Delta t > \theta_6 \\ 0, x/\Delta t \leq \theta_6 \end{cases}$, $oversd(x) = \begin{cases} x, x/\Delta t > \theta_7 \\ 0, x/\Delta t \leq \theta_7 \end{cases}$,

$flag(x) = \begin{cases} 0, x > 0 \\ 1, x = 0 \end{cases}$, $\Delta t$ is sampling time period, $\theta_6$, $\theta_7$ are *SH*-type packet number abnormality threshold; $Packet(SD_i)$ is the number of packet in $SD_i$, $i = 1, 2, ..., l$. The number of different ports in the *HF* class (*SH* class and <u>DH</u> class) of the interflow *IF* which is anomalously weighted is

$$Weight_{port} = \sum_{i=1}^{S} overp(Port(SH_i)) + \sum_{j=1}^{D} overp(Port(DH_j)) \tag{8}$$

where $overp(x) = \begin{cases} x, x/\Delta t > \theta_8 \\ 0, x/\Delta t \leq \theta_8 \end{cases}$, $\Delta t$ is sampling time period, $\theta_8$ are *SH*-type packet number abnormality threshold.

In this part we define IP Flow Multi-feature Fusion, MFF as

$$MFF_F = \frac{S + Weight_{port} + Weight_{packet}}{M + 1} \tag{9}$$

where $f(x) = \begin{cases} x, x \geq 1 \\ 1, x \leq 1 \end{cases}$.

**Definition 5.** Remember that all source semi-interactive flows *SH* are $SH_1, SH_2, ..., SH_S$. For the *SH* flows, classify the *SH* flows with the same destination IP address in the same class, and note that the number of *SH* flows with different source of IP addresses and the same destination IP address $A_i$ is $hn_i$, with the same destination IP address $A_i$. The class in which the *SH* stream resides is denoted as $HSD(hn_i, A_i)$, where $i = 1, 2, ..., l$.

Assume that all *HSD* classes are $HSD_1, HSD_2, ..., HSD_k$ and the number of different destination port numbers in the class $HSD_i$ is expressed as $Port(HSD_i)$, where $i = 1, 2, ..., K$. Then define IP Flow Address Half Interaction Anomaly Degree, HIAD as:

$$HIAD_F = \left( \sum_{i=1}^{k} \left( hn_i + weight\left(Port\left(HSD_I\right)\right)\right)\right) \tag{10}$$

where $weight(x) = \begin{cases} x, x/\Delta t > \theta_9 \\ 0, x/\Delta t \leq \theta_9 \end{cases}$, $\Delta t$ is sampling time period, $\theta_9$ is the port threshold for different destinations.

**Definition 6.**

$$SFV = \sqrt{\frac{HIAD}{FFV+1}(HIAD+FFV)} \qquad (11)$$

By analyzing the characteristics of attack flow and normal flow on the key nodes close to the attack target, the FFV reduces the interference of the normal flow for the multiple attack characteristics of the attack flow and detects attacks on key nodes close to the attack target. For the phenomenon that some of the aggregated attack flows are mixed in a large number of normal flows, the suspicious flow in the network flow is separated based on the semi-iterativeness of the attack flow source address, and the HIAD combines the source-destination IP address asymmetry of the attack flow, source address distribution, destination address concentration, and high traffic burstiness.

The SFV not only utilizes the asymmetry, distribution and concentration of the IP address of attack flow, but also takes the influencing factors of some attack flows mixed in the normal background flow into full consideration, which makes the feature expression more perfect.

**Definition 7.**

$$CDF = \frac{\sqrt{ACD+MFF}}{2} + \ln(IBF+1) \qquad (12)$$

Among them, the ACD comprehensively reflects the essential characteristics of DDoS attack flow, such as the flow burstiness, flow asymmetry and the distribution of source IP address, which is the most highly correlated with the victim-side attack flow and can directly affect the normal traffic change. The normal users and the DDoS attackers of fake source IP addresses show different characteristics in the interaction behavior of sending and receiving data packets. The nature of normal user network interaction behavior and attacker network interaction behavior in transmitting and receiving data packets is analyzed. The IBF reflects the characteristics of different network interaction behaviors between normal flow and attack flow. Based on the different behavior of the IP address and port number in the normal flow and DDoS attack flow, some essential features such as the flow asymmetry of the DDoS attack flow, the distribution of the IP address, the concentration of the attack target and sudden high traffic, the MFF can effectively separate normal and attack flows at the attack source, and better reflect the different characteristics of normal and attack flows.

The CDF combines different DDoS attack flow characteristics, such as communication asymmetry and source IP address distribution. By comparing the behavioral characteristics of normal and attack flows, we can see that the CDF completely covers the characteristics of the DDoS attack stream. Experiments show that SFV and CDF have better effects on feature extraction.

Based on the characteristics of DDoS attack traffic, GMKL detection model is applied to normal network packet traffic to detect the occurrence of DDoS attack. According to the characteristics of network communication, two feature values of SFV and CDF are proposed. The characteristics of normal traffic network traffic have certain regularity. When the change range of normal network traffic is abnormal, it can be judged that a DDoS attack has occurred.

*3.2 Model building and parameter setting*

Combining the two proposed features, the SFV and the CDF, and the GMKL we introduced, we performed parameter optimization, combined with optimized parameters, and trained the R-GMKL classifier using the feature training set to obtain the model for detection. Finally, using the test set to verify the model performance of the detection model, this completes the attack detection model optimized by the R-GMKL algorithm. Equations in display format are separated from the paragraphs of the text. Equations should be flushed to the left of the column. Equations should be made editable. Displayed equations should be numbered consecutively, using Arabic numbers in parentheses. See Eq. 1 for an example. The number should be aligned to the right margin.

On the basis of the work [Varma and Babu (2009)] and [Jain, Vishwanathan and Varma (2012)], this research is carried out. Multiple kernel learning (MKL) is defined as follows: given training set $T = \{(x_1, y_1), (x_2, y_2), (x_3, y_3) \cdots (x_n, y_n)\}$, test set $C = \{x_1^{'}, x_2^{'}, \cdots x_s^{'}\}$, where $x_i \in \mathbb{R}^d$, $x_k^{'} \in \mathbb{R}^d$, $y_i \in (-1, +1)$, $\mathbb{R}$ is a real set, $d$ as a data dimension, $i = 1, 2, \cdots, n$, $k = 1, 2, \cdots, s$. $K_1(x, x^{'}), K_2(x, x^{'}), \cdots K_M(x, x^{'})$ is the kernel function for $\mathbb{R}^d \times \mathbb{R}^d$, $\phi_1, \phi_2 \cdots \phi_M$ is the kernel map corresponding to each function. In the classic multicore learning framework SMKL, the hyperplane's objective function is:

$$f(x) = \sum_{m=1}^{M}(\omega_m, \phi_m(x)) + b \tag{13}$$

Where $\omega_m$ is the weight of each kernel function, $b$ is offset. Introducing a relaxation factor $\xi$, the objective function can be optimized to:

$$\min \psi(\omega_m, b, \xi, d) = \frac{1}{2} \sum_{m-1}^{M} \frac{1}{d_m} \|\omega_m\|_{H_m}^2 + C \sum_{i-1}^{n} \xi_i \tag{14}$$

s.t.

$$\begin{cases} y_i \sum_{m=1}^{M} \omega_m \cdot \varphi(x_i) + y_i b \geq 1 - \xi_i \\ \sum_{m=1}^{M} d_m = 1, d_m \geq 0 \\ \xi_i \geq 0 \end{cases} \tag{15}$$

Using the second-order alternating optimization, the above equation is organized into the optimization problem of the variable $d_m$,

$$\min_{d \geq 0} J(d), \sum_{m=1}^{M} d_m = 1 \tag{16}$$

s.t.

$$\begin{cases} \min_{\omega_m, b, \xi} = \frac{1}{2} \sum_{m=1}^{M} \frac{1}{d_m} \|\omega_m\|_{H_m}^2 + C \sum_{i=1}^{n} \xi_i \\ y_i \sum_{m=1}^{M} \omega_m \cdot \varphi(x_i) + y_i b \geq 1 - \xi_i \\ \xi_i \geq 0 \end{cases} \tag{17}$$

The Lagrangian function is as follows:

$$L = \frac{1}{2} \sum_{m=1}^{M} \frac{1}{d_m} \|\omega_m\|_{H_m}^2 + C \sum_{i=1}^{n} \xi_i + \sum_{i=1}^{m} \alpha_i (1 - \xi_i - y_i \sum_{m=1}^{M} \omega_m \cdot \varphi_m(x_i) + y_i b) + \sum_{i=1}^{n} v_i \xi_i \tag{18}$$

Where $\alpha_i, v_i$ is the Lagrangian operator. For the partial derivative $\omega_i, b, \xi_i$ and let the derivative be 0, the obtained extreme conditions are brought into the Lagrangian, which can be further changed to:

$$\max Q(\alpha) = -\frac{1}{2} \sum_{i,j=1}^{m} \alpha_i \alpha_j y_i y_j K_d(x_i, x_j) + \sum_{i=1}^{n} \alpha_i \tag{19}$$

s.t.

$$\begin{cases} \sum_{i}^{n} \alpha_i y_i = 0 \\ C \geq \alpha_i \geq 0 \\ K_d(x_i, x_j) = \sum_{m=1}^{M} d_m k_m(x_i, x_j) \end{cases} \tag{20}$$

The gradient descent method is used to derive the $d$ of $J(d)$, update $d$, make $d$ and $\alpha$ alternate optimization. And find an optimal solution $\alpha^* = (\alpha_1, \alpha_2, \cdots, \alpha_n)$.

That is, the original objective function eventually becomes:

$$f(x) = \sum_{i=1}^{n} \alpha_i^* y_i \sum_{m=1}^{M} d_m K_d(x_i, x_j) + b \tag{21}$$

Where $x_j \in C$. When judging the category of the test set data, simply bring the test set data into the $x_j$ above formula to determine the category corresponding to the measured data.

---

Algorithm 1: R-GMKL

---

Input: train-set (x), train-label(y)

Output: $\omega$, $b$, $d$, R, $F(x)$

Processing:

Initialization: $\omega_m$, $b_m$, $d_m$, $\xi_i$, $m=0$, $i=0$

while $m \leq M$

$\quad K \leftarrow k(d_m)$

$\quad$ Using SVM classifier and selecting a kernel function K and then can obtain $\alpha^*$

$$d_{m+1}^k = d_m^k - s_m \left( \frac{\partial r}{\partial d^k} - \frac{1}{2} \alpha^{*t} \frac{\partial H}{\partial d^k} \alpha^* \right)$$

$\quad$ Project $d_{m+1}$ onto the feasible set if any constraints are violated

$\quad m \leftarrow m+1$

$\quad F(x) = \text{sgn}(\sum_{m=1}^{M} d_m (K(x, x^T)\alpha^* + b))$

$\quad H = \omega * d$

$$R = \left| \frac{\sum_{i=1}^{n} H - H \cdot \left(\sqrt{H}\right)^T}{\sum_{i=1}^{n} H + H \cdot \left(\sqrt{H}\right)^T} \times \frac{1}{b} \right|$$

End

---

Since the goal of our algorithm is to find the best compromise between the classification accuracy of the SVM and the number of corresponding sample selection features. By iteratively selecting an individual, the higher the classification accuracy rate produced by the classifier while selecting fewer feature numbers, the higher the fitness. The fitness function we define is as follows.

$$R = \left| \frac{\sum_{i=1}^{n} H - H \cdot \left(\sqrt{H}\right)^T}{\sum_{i=1}^{n} H + H \cdot \left(\sqrt{H}\right)^T} \times \frac{1}{b} \right| \tag{22}$$

Where $H = \omega * d$; $B = b$, $\omega$, $b$ represents the weight and bias value learned in svm, and $d$ represents the weight of each kernel.

$$\min_{w,b,d} \quad \frac{1}{2} w^t w + \sum_i l(y_i, f(X_i)) + r(d) \qquad d \geq 0 \tag{23}$$

Our algorithm will continue to optimize (1) by the gradient descent algorithm, so as to continuously update the parameters $\omega$, $b$, $d$ in addition we will linearly combine the kernel function and the regularization term, and finally get the minimum value of R. The specific process is as follows:

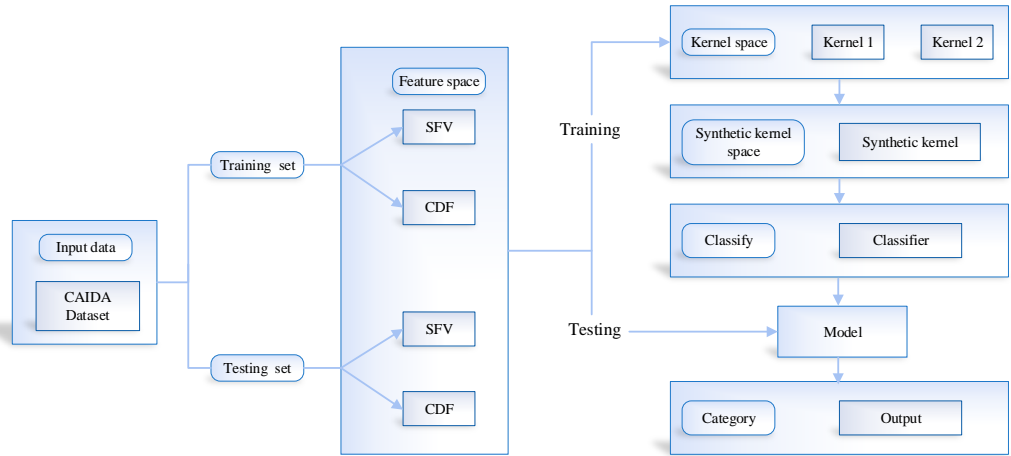

**Fig. 1:** DDoS attack detection model based on R-GMKL

The R-GMKL model actually solves the problem of two-class classification in machine learning. The detection task can be completed only by the judgment of attack. Under normal circumstances, the detection model detects that the network traffic does not have characteristic abnormality within a certain period of time. Set the detection model output flag to 1. In the case of attack, set the out-put flag of the detection model to be different from the normal condition flag, set to -1. By setting these two aspects, it can be characterized whether the network is attacked or not. With the use of the real attack detection model, it is determined whether the flag value outputted after the traffic characteristic data is abnormal after being input into the attack detection model, and the attack detection task can be completed.

**4 Experiment**

*4.1 Experimental data set and evaluation criteria*

This article uses the CAIDA "DDoS Attack 2007" dataset, which records about one hour of DDoS anonymous traffic attacks on August 4, 2007. The total size of the data set is 21 GB, which is approximately 1 hour (20:50:08 UTC - 21:56:16 UTC). In order to more reasonably evaluate the effectiveness of the experiment, we used three indicators to fully demonstrate its detection performance. The evaluation criteria used in this paper include detection rate (DR) and error rate (ER). The calculation formulas for DR and ER are as follows:

$$\begin{cases} DR = \dfrac{TN}{TN+FN} \\ ER = \dfrac{FN+FP}{TP+FP+TN+FN} \end{cases} \quad (25)$$

Where TP indicates that the number of normal test samples that are correctly marked, FP indicates the number of normal test samples that are incorrectly marked, TN indicates the number of attack test samples that are correctly marked, and FN indicates the number of attack test samples that are incorrectly marked. The experiment uses the above three evaluation criteria to compare with the original methods, the SVM method and the SMKL method to verify the effectiveness of the algorithm.

### *4.2 Experimental results and analysis*

We did three sets of experiments. The first group is in the normal network flow, the second group is when a large number of background streams are mixed with a small number of attack streams, and the third group is when a large number of attack streams are mixed with a small number of normal streams. The basis of the experiment is to obtain the positive and negative sample sets by extracting the feature data of the attack stream and the normal stream.

Fig. 2 is a comparison of the network data stream and the feature values extracted herein. Fig. 2(a) shows the number of network stream packets with an acquisition time of 116s and a sampling time of 1s. Fig. 2(b) and Fig. 2(c) respectively sample the normal stream and the attack stream and calculate the SFV and CDF time series. The feature extraction period is set to 1s, and a total of 116 positive samples and 116 negative samples are extracted.

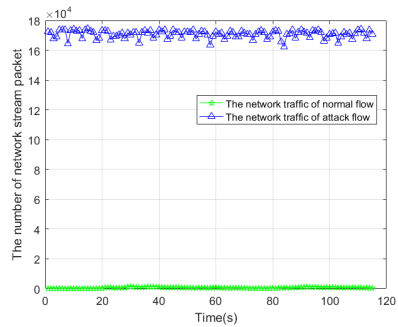

**Fig. 2(a):** The network traffic of normal flow and attack flow

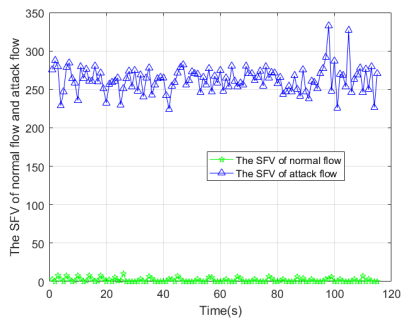 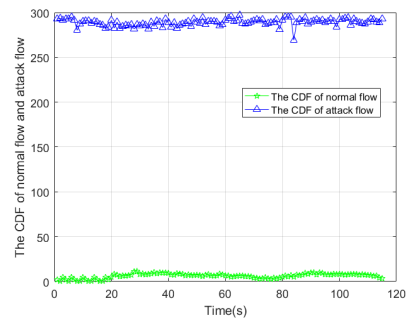

**Fig. 2(b):** The SFV of normal flow and attack flow    **Fig. 2(c):** The CDF of normal flow and attack flow

**Fig. 2:** The comparison of network traffic and eigenvalues between normal flow and attack flow

Fig. 2(a) shows that if the number of data packets is only counted for the network stream,

the value is too large, which leads to an increase in computer memory consumption when calculating the normal flow and the attack flow. As can be seen from Fig. 2(b), the data of the normal stream is always in a stable state. After early data, this feature can greatly differentiate between normal traffic and abnormal traffic, and more greatly influence the classifier to make better decisions. The data of the attack stream is mainly between 200 and 300, and the resolution is strong. As shown in Fig. 2(c), the attack stream feature values are concentrated between 250 and 300, and the normal stream data fluctuates within a small range. However, this does not affect the model's classification of normal and attack streams. Using SFV and CDF as features can significantly reflect the difference between attack traffic and normal traffic. In addition, in general, the data of the attack stream is evenly distributed, and the normalization effect is good in further work because the extremum is small and easy to standardize.

Through analysis, we find that the SFV and CDF features are interrelated and complementary. They can show great differences between normal flow and attack flow, and correctly classify normal flow data and attack flow data.

*4.2.1 Eigenvalues in early attack*

In order to better illustrate the robustness of R-GMKL, our simulation has done the following three kinds of cases. The first case is early attack, the second is impulse attack, and the third case is intermittent attack.

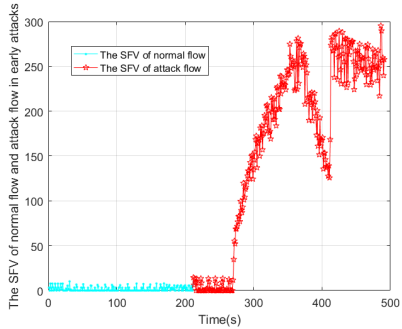 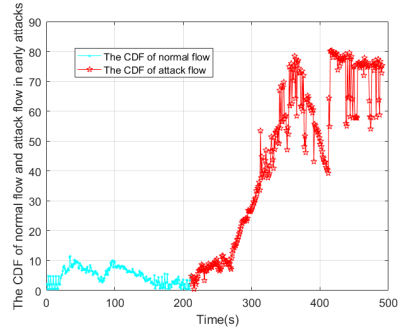

**Fig. 3(a):** The SFV of normal flow and attack flow in early attacks  **Fig. 3(b):** The CDF of normal flow and attack flow in early attacks

**Fig. 3:** The eigenvalues of normal flow and attack flow in early attacks

In this part of the experiment, the captured samples are 491 data samples with data concentration in the early stage of the attack (the attack behavior was detected at the beginning), including 211 normal sample data and 280 attack sample data. Select 344 sample points as the training set and 147 sample points as the test set.

As shown in Fig. 3(a), by computing SFV, our algorithm can detect attack flow signatures at an early stage. Although it does not have the same degree of discrimination as the CDF at the beginning, the eigenvalues of the attack are particularly concentrated, which is a good reflection of the difference between the normal sample and the attack sample. As shown in Fig. 3(b), by calculating the CDF, our algorithm can also detect early attack stream samples. The normal stream samples and attack stream samples have strong discrimination, which is obviously different from the normal stream characteristics. Accuracy is used as an evaluation index. By comparing different kernel function parameters and R values under regularization parameters, we can select a combination with higher precision and retain parameters for comparison experiments.

**Table 1:** The comparison of parameter selection results.

| Group | 1 | 2 | 3 | 4 |
|---|---|---|---|---|
| Kernel function parameter | Product of RBF kernels | Sum of RBF kernels | Product of RBF kernels | Sum of RBF kernels |
| Regularization paradigm | L1 | L1 | L2 | L2 |
| R | 0.49 | 0.71 | 0.84 | 1.08 |
| Accuracy (%) | 93.2% | 89.8% | 88.4% | 87.1% |

From Tab. 1, in the general network flow, the combination of the regularization parameter selection L1 and the kernel function Product of RBF kernels is selected, the R value is the smallest, and the accuracy is high, so in the general network flow, we select Product of the RBF kernels for the next experiment, as well as the L1 regularization parameters. In order to verify the superiority and reliability of R-GMKL algorithm in detecting DDoS attacks, classical methods such as GMKL, SMKL and SVM are compared, and the correctness of kernel function and regularization parameter selection methods are verified from DR and ER. The comparison results are shown in Tab. 2.

**Table 2:** The comparison of attack detection algorithm.

|  | Simple MKL | SVM | R-GMKL |
|---|---|---|---|
| DR(%) | 79.8% | 59.5% | 88.1% |
| ER(%) | 11.6% | 23.1% | 6.8% |

From the comparison of the results in Tab. 2, it is found that the detection rate and the error rate of the DDoS attack detection method based on the R parameter optimization GMKL are 96.4% and 0.02%. Compared with the popular SVM and SMKL algorithms, it has higher detection rate and lower error rate. It can be concluded that the DDoS attack detection method based on R parameter optimization GMKL proposed in this paper can effectively improve the detection rate of DDoS attacks and reduce the error rate. It has better comprehensive performance, can effectively identify DDoS attacks, and provides new DDoS attack detection methods and means.

*4.2.2 Eigenvalues in impulse attack*

In this part of the experiment, a situation is simulated that a large number of normal flows are mixed with a small number of attack flows based on the sample points of the data set. There are 491 data samples, including 384 normal samples and 107 attack samples. According to the selection method of Chapter 4.2.1, 344 samples were selected as training set and 147 samples were selected as test set.

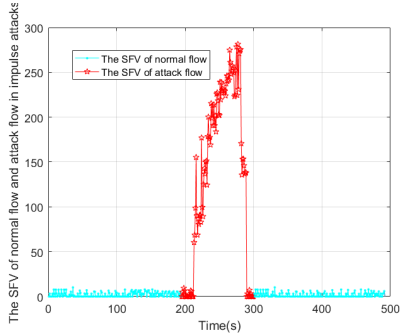 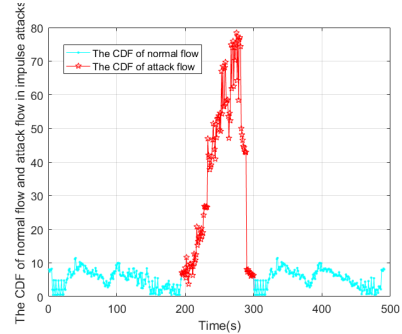

**Fig. 4(a):** The SFV of normal flow and attack flow in impulse attacks

**Fig. 4(b):** The CDF of normal flow and attack flow in impulse attacks

**Fig. 4:** The eigenvalues of normal flow and attack flow in impulse attacks

As shown in Fig. 4(a) and Fig. 4(b), we inject a small amount of attack streams into a large number of normal streams at the same time in order to distinguish between normal and attack streams in time. As shown in Fig. 4(a), the SFV characteristic enables the attack stream characteristic values to aggregate together, and the attack part tends to be more stable. Through the analysis of the data, this feature can distinguish the normal flow from the abnormal flow and have a great influence on the decision-making of the classifier. As shown in Fig. 4(b), normal traffic and attack traffic can be well separated by CDF features. As long as abnormal traffic is detected, CDF can accurately distinguish it from normal traffic characteristic values.

**Table 3:** The comparison of parameter selection results.

| Group | 1 | 2 | 3 | 4 |
| --- | --- | --- | --- | --- |
| Kernel function parameter | Product of RBF kernels | Sum of RBF kernels | Product of RBF kernels | Sum of RBF kernels |
| Regularization paradigm | L1 | L1 | L2 | L2 |
| R | 1.10 | 1.62 | 2.98 | 1.92 |
| Accuracy (%) | 96.6% | 96.6% | 91.84% | 91.84% |

Tab. 3 describes the mixing of a small number of attack streams in a normal stream. Through observation, it can be found that the accuracy of the first group with the lowest R value is the highest after comparing four groups of experiments. L1 regularization was applied in the first two groups, and the accuracy of the two groups was the highest. It can be concluded that the generalization ability of L1 regularization is much better than that of L2 regularization when a large number of normal flows are mixed with a small number of attack flows. Therefore, we choose the combination of the first set of parameters for the next experiment.

**Table 4:** The comparison of attack detection algorithm.

|  | Simple MKL | SVM | R-GMKL |
| --- | --- | --- | --- |
| DR(%) | 62.5% | 59.4% | 84.4% |
| ER(%) | 8.2% | 8.8% | 3.4% |

It can be observed from Tab. 4 that when a large number of normal flows are mixed with a small number of attack flows, the overall performance of the algorithm is R-GMKL, SMKL and SVM in order from high to low according to three evaluation criteria. As high as 84.4% detection rate can fully prove that R-GMKL can detect impulse attacks to some extent.

*4.2.3 Eigenvalues in intermittent attack*

Through the analysis of attack types and characteristics, we can know that some attacks are regularly presented, while some attacks are irregular, such as intermittent attacks caused by offensive and defensive confrontation. In one case, in order to avoid detection, a small attack will be launched first to see if it can be detected, and then continue to attack. Another case is that, because an attack is detected, it will immediately take another way to attack, switching between different attacks. In this part of the experiment, based on the sample points of the data set, there are 491 data samples, including 80 normal samples and 411 attack samples. The selection method also uses 70% as training set and 30% as testing set. 344 samples are selected as training set and 147 samples are selected as testing set.

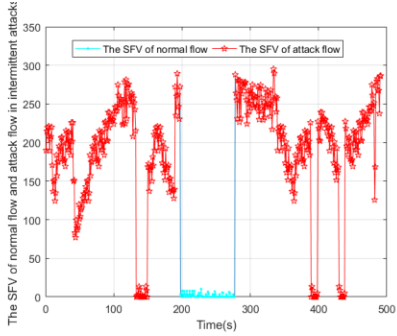 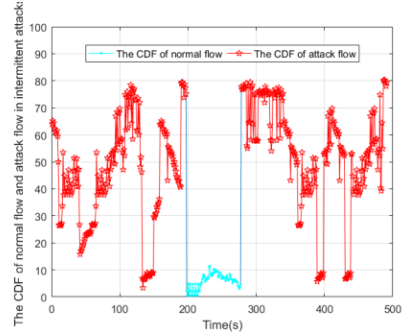

**Fig. 5(a):** The SFV of normal flow and attack flow in intermittent attacks

**Fig. 5(b):** The CDF of normal flow and attack flow in intermittent attacks

**Fig. 5**: The eigenvalues of normal flow and attack flow in intermittent attacks

As shown in Fig. 5(a) and Fig. 5(b), the discrimination of SFV features is greater than that of CDF at the same attack data points. In the attack flow after normal flow, these two characteristics are concentrated and the representation ability is stable. CDF eigenvalues fluctuate greatly. Generally speaking, when a large number of attacks occur, both eigenvalues can detect the time point of attack in time.

**Table 5:** The comparison of parameter selection results.

| Group | 1 | 2 | 3 | 4 |
| --- | --- | --- | --- | --- |
| Kernel function parameter | Product of RBF kernels | Sum of RBF kernels | Product of RBF kernels | Sum of RBF kernels |
| Regularization paradigm | L1 | L1 | L2 | L2 |
| R | 0.95 | 0.98 | 0.96 | 0.97 |
| Accuracy (%) | 88.44% | 87.76% | 87.76% | 87.76% |

Tab. 5 shows the case where a small amount of normal flow is trapped in the attack stream. By observation, the smallest R value is the first group, and its corresponding accuracy is also the highest. Therefore, in this case, we choose the Product of RBF kernels and the L1 regularization parameters for the next experiment.

**Table 6:** The comparison of attack detection algorithm.

|        | Simple MKL | SVM   | R-GMKL |
|--------|------------|-------|--------|
| DR(%)  | 78.9%      | 78.9% | 86.2%  |
| ER(%)  | 17.7%      | 17.7% | 11.6%  |

As can be seen from Tab. 6, when a large number of attack streams are mixed with a small amount of normal flow background, the detection ability of SVM and SMKL are equal, and the detection rate of R-GMKL is higher than that of the former two, and the error rate is lower.

In summary, on the basis of data set, this paper simulates three kind cases of attacks: early attack, impulse attack and intermittent attack. R-GMKL method has the highest detection rate and the lowest error rate. The experimental results show that the smaller the R value, the higher the accuracy, which provides a scientific method for the selection of the kernel function and regularization parameters. In addition, the proposed method based on R parameter optimization GMKL is proved to be robust.

## 5 Conclusion

Aiming at the problems of single DDoS attack detection method, such as single attack type, high false negative rate and false positive rate, we propose a DDoS attack detection method based on GMKL. Two characteristics of SFV and CDF for describing the characteristics of network flow are defined. Based on these two features, a DDoS attack detection model based on R parameter optimization GMKL is established. A method for kernel function and regularization parameter selection is proposed, which reduces the human-induced error to a certain extent and provides a scientific basis. The experimental results show that the proposed method can detect DDoS attacks effectively and early, and have higher detection rate and lower false negative rate than similar methods.

**Acknowledgement**: This work was supported by the Hainan Provincial Natural Science Foundation of China [2018CXTD333,617048]; National Natural Science Foundation of China [61762033, 61702539]; Hainan University Doctor Start Fund Project [kyqd1328]; Hainan University Youth Fund Project [qnjj1444].